\UseRawInputEncoding

\documentclass[pdflatex,sn-mathphys-num]{sn-jnl}

\usepackage{graphicx}%
\usepackage{multirow}%
\usepackage{amsmath,amssymb,amsfonts}%
\usepackage{amsthm}%
\usepackage{mathrsfs}%
\usepackage[title]{appendix}%
\usepackage{xcolor}%
\usepackage{textcomp}%
\usepackage{manyfoot}%
\usepackage{booktabs}%
\usepackage{algorithm}%
\usepackage{algorithmicx}%
\usepackage{algpseudocode}%
\usepackage{listings}%

\theoremstyle{thmstyleone}%

\theoremstyle{thmstyletwo}%

\theoremstyle{thmstylethree}%

\begin{document}

\title[Article Title]{Exact linear representations of the nonlinear
Cauchy problems and their smooth solutions}

\author {\fnm{Yurii \,N.} \sur{Kosovtsov}}\email{yunkosovtsov@gmail.com}

\affil{\city{Lviv}, \country{Ukraine}}

\abstract {This paper introduces a novel approach to the exact linearization of nonlinear differential equations by constructing their linear representations.
We establish conditions under which such representations exist for Cauchy problems involving nonlinear partial differential equations.
The method introduces a linear operator $A$, which reveals an equivalence between the original nonlinear equation and two other equations, one of which is linear.
Crucially, all three equations share a common formal operator solution given by the Taylor series expansion of the sought function $v(t,x)$.
Based on the Borel-Whitney lemma, we prove that if the parameters of operator $A$ are smooth, this formal series corresponds to a genuine smooth function, $\tilde{v}(t,x)$,
that solves all three equivalent equations. For systems of equations (both ordinary and partial), we propose a linearization method based on embedding the nonlinear system
into a higher-dimensional, specially structured linear system. This technique allows us to construct general solutions in explicit form for a broad class of nonlinear DEs.
The effectiveness of the approach is demonstrated by finding smooth solutions for several physically significant problems. The Navier-Stokes and Euler equations are also considered
as non-trivial applications of the theory.}


\maketitle

\section{Equivalence of Equations}

Finding exact solutions to nonlinear differential equations is one of the fundamental problems of modern mathematical physics. Of particular interest is the possibility of linearizing such equations, i.e., representing them as linear equations. One of the motivations for this work is the formal scheme presented in \cite{Kosovtsov} for obtaining solutions to nonlinear equations using an operator method.

The purpose of this section is to obtain \emph{formal} solutions $v=v(t,x)$ for nonlinear equations of the form ($t,x\in \mathbb{R}\times \mathbb{R}^n$):
\begin{equation}
\frac{\partial v}{\partial t} = F(x,v,\textbf{D}_j^{\alpha_j}v), \qquad v(t,x)|_{t=0}=u(x),
\label{NLdE2}
\end{equation}
where $\textbf{D}_j^{\alpha_j}v$ represents a given finite sequence of derivatives of the function $v$ with respect to the components of $x \in \mathbb{R}^n$.

Assuming that the functions $u$ and $F$ are smooth (of class $C^\infty$) in all their variables, we define, in the spirit of \cite{Kosovtsov}, the differential operator
\begin{equation}
 A=\int_{\mathbb{R}^n} d\zeta \, F(\zeta,u(\zeta),\textbf{D}_j^{\alpha_j}u(\zeta))\frac{\delta}{\delta u(\zeta)},
\label{opA}
\end{equation}
where the function $F(x,u(x),\textbf{D}_j^{\alpha_j}u(x)) = F(x,u_0,u_1,...,u_m)$ is \emph{the right-hand side of equation} (\ref{NLdE2}) \emph{at} $t=0$. Since $u$ is assumed to be an \emph{arbitrary but fixed} function, $F$ can be considered a linear multiplication operator, and $\frac{\delta}{\delta u(\zeta)}$ is the functional derivative. The symbol $d\zeta$ is placed at the beginning of the integral to emphasize that the operator $\frac{\delta}{\delta u(\zeta)}$ acts first, after which the integration is performed.

Here, the functional derivative is a linear mapping with the property
\[\frac{\delta u(x)}{\delta u(\zeta)} = \delta(x-\zeta),\]
where $\delta(x-\zeta)$ is the Dirac delta function. The chain rule is also valid in this context.

We emphasize that this operator $A$ \emph{does not depend on time $t$}.

The operator $A$ is linear, satisfying the following properties:
\[A [c G(u(x))] = c A G(u(x)), \qquad \mathrm{for\; any\; scalar}\; c\in \mathbb{R}, \]
\[A [G(u(x))+H(u(x))] = A G(u(x))+A H(u(x)), \qquad \mathrm{for\; any}\; G,H\in C^\infty.\]
Furthermore, operator $A$ is a \emph{derivation}, since
\[A[G(u(x)) \cdot H(u(x))] = [AG(u(x))] \cdot H(u(x)) + G(u(x)) \cdot A H(u(x)),\]
and in the particular case of acting on $u(x)$, we have
\[Au(x)=F(x,u(x),\textbf{D}_j^{\alpha_j}u(x)).\]
Note also that $A^j u \in C^\infty$ for any $j \in \mathbb{N}$.

First, let us consider the associated \emph{linear operator} differential equation:
\begin{equation}
\frac{\partial v(t,x)}{\partial t} = Av(t,x), \qquad v(x,t)|_{t=0}=u(x).
\label{LdE2}
\end{equation}
Transforming this into the equivalent integral equation
\begin{equation}
v(t,x) = u(x)+\int_0^t Av(\tau,x) d\tau
\label{LdI2}
\end{equation}
and solving by the Picard iteration method, starting with $v_0(t,x)=u(x)$, we have:
\[v_1(t,x)=u(x)+\int_0^t A u(x) d\tau = u(x)+tAu(x);\]
\[v_2(t,x)=u(x)+\int_0^t A[u(x)+\tau Au(x)] d\tau = u(x)+tAu(x)+\frac{t^2}{2}A^2u(x);\]

\[...\]

\begin{equation}
v(t,x)=\sum_{n=0}^\infty\frac{t^n A^n}{n!}u(x)=e^{tA}u(x).
\label{Ldsol2}
\end{equation}
Thus, we obtain a solution to equation (\ref{LdE2}) in the form of a formal functional series:
\begin{equation}
v(t,x)=\sum_{n=0}^\infty\frac{t^n a_n(x)}{n!},
\label{FS}
\end{equation}
where $a_n(x)=A^n u(x)$. It is natural to assume that (\ref{Ldsol2}) \emph{is the formal Taylor series expansion of the function $v(t,x)$} around $t=0$, and therefore that
\begin{equation}
a_n(x)=A^n u(x) = \frac{\partial^n v(t,x)}{\partial t^n}\bigg|_{t=0}.
\label{LdS2}
\end{equation}
This holds true (with some reservations), since using equation (\ref{LdE2}) yields the following chain of equalities:
\[\frac{\partial v(t,x)}{\partial t}\bigg|_{t=0} = Av(t,x)|_{t=0} = Au(x),\]
\[\frac{\partial^2 v(t,x)}{\partial t^2}\bigg|_{t=0} = A\frac{\partial v(t,x)}{\partial t}\bigg|_{t=0} = A^2u(x),\]
\[\vdots\]
\[\frac{\partial^n v(t,x)}{\partial t^n}\bigg|_{t=0} = A^n u(x).\]

We now note another important point. Differentiating the operator solution (\ref{Ldsol2}) with respect to $t$ and noting that $A$ commutes with $e^{tA}$, we have
\begin{equation}
\frac{\partial v(t,x)}{\partial t} = A e^{tA}u(x) = e^{tA}Au(x).
\label{Twoeq}
\end{equation}
As a result, we obtain \emph{three formally equivalent} equations. The first is the original linear equation (\ref{LdE2}):
\[\frac{\partial v(t,x)}{\partial t} = A(e^{tA}u(x)) = Av(t,x).\]
The second equation is
\[\frac{\partial v(t,x)}{\partial t} = e^{tA}Au(x).\]
Under certain conditions, this equation transforms into the nonlinear equation (\ref{NLdE2}):
\[\frac{\partial v(t,x)}{\partial t}=F(x,v,\textbf{D}_j^{\alpha_j}v), \qquad v(t,x)|_{t=0}=u(x).\]
To show this, we must consider the properties of the operator $e^{tA}$. Note that $e^{-tA}$ is formally the inverse of $e^{tA}$, since $e^{-tA}e^{tA}=I$.

Let $g(u(x))$ be a multiplication operator. As can be seen from the definition of the derivation operator $A$, the commutator $[g(u(x)),A]$ is also a multiplication operator (i.e., a function).
Now consider the operator construction
\[B=e^{tA} g(u(x)) e^{-tA}.\]
Since the commutator $[g(u(x)),A]$ is a function, all repeated commutators are also functions. Thus, using the well-known Baker-Campbell-Hausdorff formula (see (26)-(28) in \cite{Kosovtsov3}), we conclude that the operator $B$ is also a function, and
\begin{equation}
B = (e^{tA}g(u(x))), \label{hom}
\end{equation}
where the outer brackets on the right-hand side emphasize that the action of $e^{tA}$ is restricted to $g(u(x))$. An analogous conclusion holds for $e^{-tA}g(u(x))e^{tA}$.

Let us consider the following chain of identities:
\begin{align}
e^{tA} g_1 g_2 \dots g_p &= e^{tA}g_1 e^{-tA} e^{tA}g_2 e^{-tA} \dots e^{tA}g_p \notag
\\&= (e^{tA} g_1) (e^{tA} g_2) \dots (e^{tA} g_p). \notag
\end{align}
This observation, combined with the linearity of operator $A$, implies that for any smooth function $F(g_1, g_2, \dots, g_p)$ that can be expanded in a formal power series with respect to its arguments, we obtain the following important property:
\begin{equation}
e^{tA} F(g_1, g_2, \dots, g_p) = F(e^{tA}g_1, e^{tA}g_2, \dots, e^{tA}g_p). \label{homom}
\end{equation}
This property of $e^{tA}$ is analogous to the familiar property of the shift operator $\exp\{t\frac{\partial}{\partial s}\}$, as the shift operator is a particular case of an operator exponent.

Henceforth, we will assume that $F$ is a smooth function with respect to all its arguments. Then, as a consequence of (\ref{homom}) and taking into account (\ref{Ldsol2}), we have
\begin{equation}
e^{tA}F(x,u(x),\textbf{D}_j^{\alpha_j}u(x))=F(x,e^{tA}u(x),\textbf{D}_j^{\alpha_j}e^{tA}u(x))=F(x,v,\textbf{D}_j^{\alpha_j}v).
\label{Twoeq2}
\end{equation}
This implies that the second \emph{linear equation}, $\frac{\partial v}{\partial t} = e^{tA}Au(x)$, is \emph{equivalent to the nonlinear equation} (\ref{NLdE2}).

From a combination of these equations, one can obtain the third equivalent equation:
\[Av=F(x,v,\textbf{D}_j^{\alpha_j}v), \qquad v(t,x)|_{t=0}=u(x).\]
The equivalence of these equations implies that \emph{all three equations} share the same formal solution (\ref{Ldsol2}):
\[v(t,x)=\sum_{n=0}^\infty\frac{t^n A^n}{n!}u(x)=e^{tA}u(x),\]
for which
\[\frac{\partial^n v(t,x)}{\partial t^n}\bigg|_{t=0}=A^n u(x).\]
This last equality can be verified directly (albeit via cumbersome calculations) for specific examples of nonlinear equations of type (\ref{NLdE2}) with an operator $A$ defined according to (\ref{opA}). Both sides of the partial differential equation can be differentiated $k$ times with respect to $t$ to yield recurrence relations for the coefficients of the formal power series for $v(t,x)$, which determine $\frac{\partial^n v(t,x)}{\partial t^n}|_{t=0}$. Comparing these coefficients with $A^k u(x)$, we find they are identical.

Finally, noting that the formal Taylor series expansion can be represented using the differential shift operator $e^{t\frac{\partial }{\partial s}}$ (where $\frac{\partial }{\partial s}$ is a \emph{derivation}) as
\[v(t,x)=\sum_{n=0}^\infty\frac{t^n a_n(x)}{n!}=\left[e^{t\frac{\partial }{\partial s}}v(s,x)\right]_{s=0},\]
we can write
\begin{equation}
\left[e^{t\frac{\partial }{\partial s}}v(s,x)\right]_{s=0}=e^{tA}u(x).
\label{EXA}
\end{equation}
The shift operator has the well-known property that for any differentiable function $f(s)$,\footnotemark
\footnotetext{The shift operator we use is related to the solution of the linear partial differential equation $\frac{\partial g(t,s)}{\partial t} = \frac{\partial g(t,s)}{\partial s}$ with initial condition $g(t,s)|_{t=0}=f(s)$. Using the Picard iteration method with the operator $A=\frac{\partial }{\partial s}$ and starting from $g_0(t,s)=f(s)$, we obtain the solution $g(t,s)=e^{t\frac{\partial }{\partial s}}f(s)$. While this requires $f(s)$ to be infinitely differentiable, the equation is known to have the solution $g(t,s)=f(t+s)$, which only requires $f(s)$ to be differentiable once. The property $e^{t\frac{\partial }{\partial s}}f(s)= f(t+s)$ is what gives the operator its name.}
\[e^{t\frac{\partial }{\partial s}}f(s)=f(e^{t\frac{\partial }{\partial s}}s)=f(s+t),\]
that is, the operator $e^{t\frac{\partial }{\partial s}}$ can be brought "inside" a differentiable function. The same holds for a function $f(s,x)$ of several variables that is differentiable with respect to $s$.

Let us suppose that
\begin{equation}
v(t,x)=\left[e^{t\frac{\partial }{\partial s}}v(s,x)\right]_{s=0}=e^{tA}u(x).
\label{supp}
\end{equation}
Then
\[v(t+s,x)=e^{t\frac{\partial }{\partial s}}v(s,x)=e^{(t+s)A}u(x).\]
Differentiating the last expressions, we obtain
\[\frac{\partial v(t+s,x)}{\partial t}=\frac{\partial v(t+s,x)}{\partial s}=\frac{\partial}{\partial s}\left(e^{t\frac{\partial }{\partial s}}v(s,x)\right)=\frac{\partial}{\partial s}\left(e^{(t+s)A}u(x)\right)=Ae^{(t+s)A}u(x).\]
From this chain, we need the equality
\begin{equation}
\frac{\partial}{\partial s}\left(e^{t\frac{\partial }{\partial s}}v(s,x)\right)=Ae^{(t+s)A}u(x).
\label{difshift}
\end{equation}
We now expand both sides of the \emph{nonlinear} equation (\ref{NLdE2}) into a Taylor series by applying the shift operator $e^{t\frac{\partial }{\partial s}}$ and evaluating at $s=0$:
\[\left[e^{t\frac{\partial }{\partial s}}\left(\frac{\partial v(s,x)}{\partial s}\right)\right]_{s=0}=\left[e^{t\frac{\partial }{\partial s}}F(x,v(s,x),\textbf{D}_j^{\alpha_j}v(s,x))\right]_{s=0}.\]
Using the properties of the shift operator, this becomes
\[\frac{\partial}{\partial t}\left[e^{t\frac{\partial }{\partial s}}v(s,x)\right]_{s=0}=F\left(x,\left[e^{t\frac{\partial }{\partial s}}v(s,x)\right]_{s=0},\textbf{D}_j^{\alpha_j}\left[e^{t\frac{\partial }{\partial s}}v(s,x)\right]_{s=0}\right).\]
Recognizing that $\frac{\partial}{\partial t} = \frac{\partial}{\partial s}$ for the function $v(t+s,x)$, we can write
\[\frac{\partial}{\partial s}\left[e^{t\frac{\partial }{\partial s}}v(s,x)\right]_{s=0}=F\left(x,\left[e^{t\frac{\partial }{\partial s}}v(s,x)\right]_{s=0},\textbf{D}_j^{\alpha_j}\left[e^{t\frac{\partial }{\partial s}}v(s,x)\right]_{s=0}\right).\]
Using (\ref{EXA}) and (\ref{difshift}), we arrive at
\[Ae^{tA}u(x)=F(x,e^{tA}u(x),\textbf{D}_j^{\alpha_j}e^{tA}u(x)).\]
Denoting $v(t,x)=e^{tA}u(x)$, we obtain the equation equivalent to (\ref{NLdE2}):
\begin{equation}
Av(t,x)=F(x,v(t,x),\textbf{D}_j^{\alpha_j}v(t,x)), \qquad v(t,x)|_{t=0}=u(x).
\label{Nonl3}
\end{equation}
Combining (\ref{NLdE2}) and (\ref{Nonl3}) yields one more equivalent equation:
\begin{equation}
\frac{\partial v(t,x)}{\partial t}=Av(t,x), \qquad v(t,x)|_{t=0}=u(x),
\notag
\end{equation}
which is precisely the linear equation (\ref{LdE2}) for which we found the formal solution (\ref{Ldsol2}), consistent with our assumption (\ref{supp}).

Thus, we can conclude:

\textbf{Proposition 1.} \emph{Let the function} $F(x,v_0,v_1,...,v_m)$ \emph{in equation} (\ref{NLdE2}) \emph{be smooth (of class} $C^\infty$) \emph{with respect to all its variables, and let the operator $A$ be defined by} (\ref{opA}). \emph{Then the equations} (\ref{NLdE2}), (\ref{LdE2}), \emph{and} (\ref{Nonl3}) \emph{are formally equivalent and share the same solution} (\ref{Ldsol2}).

Note that Proposition 1 deals with a nonlinear PDE (\ref{NLdE2}), a linear operator equation (\ref{LdE2}), and a nonlinear equation with \emph{variational} derivatives (\ref{Nonl3}). All these equations are linked by the operator $A$, which is constructed from the original nonlinear equation (\ref{NLdE2}). This result has parallels with well-known exact representations of \emph{nonlinear} ODEs by \emph{linear} PDEs. For nonlinear PDEs in a broader sense, this result was first suggested, albeit as a hypothesis, in \cite{Kosovtsov} (see also \cite{Kosovtsov2}).

\emph{Remark.} The expression (\ref{LdS2}) unambiguously relates the action of the known operator $A$ on the known (fixed) function $u(x)$ to the action of the differentiation operator $\frac{\partial}{\partial t}$ on the unknown function $v(t,x)$ at $t=0$. From a practical standpoint, (\ref{LdS2}) implies that if one could assume the convergence of the Taylor series for $v(t,x)$, one could then justify the convergence of the series $\sum_{n=0}^\infty \frac{t^n A^n}{n!}u(x)$. This is typically not feasible, as $v(t,x)$ is unknown.

Conversely, the formal operator solution involves only known functions: the arbitrary but fixed function $u(x)$ and the function $F$, which generates the set of functions $a_n=A^n u(x)$. Therefore, it may be more realistic to use a version of an "inverse" Taylor theorem to establish properties of $v(t,x)$ based on the known properties of $u$, $F$, and the resulting coefficients $a_n$.

Moreover, the recovery of a function $v(t,x)$ from its time derivatives $\frac{\partial^k v}{\partial t^k}|_{t=0}$ is only possible up to a flat function and is therefore not unique. The recovery of $v(t,x)$ from the coefficients $A^k u$, however, is unique by construction.

Here is a quote from \cite{Poston}:
"Normally, in applied mathematics or physics, the Taylor series has been used to approximate $f$, and it is not surprising that analyticity has been emphasized. But in fact analyticity is neither necessary nor sufficient for such approximations to be valid, in terms of the use to which they are put. Analyticity has been over-emphasized, and the vanishing of the remainder term $R_k = f(x_0, x) - (a_0 + a_1 x + \dots + a_kx^k)$ as $k$ tends to $\infty$ has figured too prominently, thus - in an elegant phrase of Zeeman's - allowing the tail of the Taylor series to wag the dog."

\section{The Borel-Whitney Lemma}

The Borel-Whitney lemma states that any smooth function in a neighborhood of a point can be represented by its formal Taylor series up to a flat function—a term whose own Taylor series at that point is identically zero. This fact is useful for proving various assertions about the local properties of functions, mappings, and PDE solutions, among others.

This line of research, initiated by E. Borel \cite{Borel} and continued by H. Whitney \cite{Whitney1}, \cite{Whitney2}, continues to be developed by many authors. The current state of this problem can be found in \cite{Brudnyi} and in a series of papers by Ch. Fefferman (see, for example, \cite{Fefferman}).

Consider a formal power series in variables $t=(t_1,...,t_p)$ whose coefficients are smooth functions of the variables $x=(x_1,...,x_m)$:
\begin{equation}
\sum_{|\alpha|=0}^\infty b_{\alpha}(x)t^\alpha.
\label{series}
\end{equation}

A natural question arises: does every formal series (\ref{series}) correspond to a smooth function $V(t,x)$ whose Taylor series in $t$ at $t=0$ coincides with the given series? The answer to this question is affirmative.

For the purposes of this paper, we only need the elementary lemma attributed to Emile Borel and Hassler Whitney \cite{Borel}-\cite{Whitney2}, \cite{Golubitsky}, \cite{Hormander}.

\textbf{Lemma (Borel-Whitney).}
\emph{For any formal series of the form} (\ref{series}) \emph{and any open bounded domain} $\Omega \subset \mathbb{R}^m$, \emph{there exists a smooth function} $V(t,x): \mathbb{R}^p \times \Omega \rightarrow \mathbb{R}$ \emph{whose Taylor series in the variable $t$ at the point} $t=0$ \emph{coincides with} (\ref{series}).

\emph{Proof.} For simplicity, we assume that $p=1$ and $m=1$; in the general case, the proof is analogous, but the formulas are more cumbersome.

Let $\psi : \mathbb{R} \rightarrow \mathbb{R}$ be a smooth function (a cutoff function) such that
\[ \psi(t) =
\begin{cases}
1, & |t|\leq 1/2, \\
0, & |t|\geq 1.
\end{cases}
\]

We will show that for any formal series
\begin{equation}
\sum_{n=0}^\infty b_n(x) t^n, \qquad t,x \in \mathbb{R}
\label{series2}
\end{equation}
with smooth coefficients $b_n(x)$ and for any open bounded domain $\Omega \subset \mathbb{R}$, there exists a smooth function $V(t,x): \mathbb{R} \times \Omega \rightarrow \mathbb{R}$ whose Taylor series in the variable $t$ at $t=0$ coincides with (\ref{series2}).

We seek the function $V(t,x)$ in the form of the series
\begin{equation}
V(t,x)=\sum_{n=0}^\infty b_n(x) t^n \psi(t/r_n),
\label{cutF}
\end{equation}
where $b_n(x)$ are the coefficients from (\ref{series2}) and the numbers $r_n>0$ are defined by
\[r_n=\frac{1}{n!(1+\beta_n)},\quad \mathrm{where}\quad \beta_n=\max_{0 \le i \le n}\,\max_{x\in\overline{\Omega} } |b_n^{(i)}(x)|,\]
where $\overline{\Omega}$ is the closure of $\Omega$, and $b_n^{(i)}(x)$ is the $i$-th derivative of $b_n(x)$ with respect to $x$.

It is not difficult to verify that for any integer $k \ge 0$, the following estimate holds:
\begin{align}
 \sum_{n=k+1}^\infty \sup_{x \in \overline{\Omega}} |D_x^k (b_n(x) t^n\psi(t/r_n))|& \leq \sum_{n=k+1}^\infty \beta_n r_n^n = \sum_{n=k+1}^\infty \frac{\beta_n}{(n!(1+\beta_n))^n} \notag \\\leq
& \sum_{n=k+1}^\infty \frac{1}{(n!)^n (1+\beta_n)^{n-1}} < \infty.
\end{align}
This shows that the series obtained by differentiating the terms of (\ref{cutF}) with respect to $x$ converges absolutely and uniformly on $\mathbb{R} \times \overline{\Omega}$. This implies that the series (\ref{cutF}) defines a function $V(t,x)$ that is $C^\infty$ with respect to $x$.

Similarly, one can show that the series obtained by differentiating the terms of (\ref{cutF}) $k$ times with respect to $t$ and $j$ times with respect to $x$ also converges absolutely and uniformly. For instance, for a single time derivative, we have
\[\left| \frac{d}{dt}(t^n\psi(t/r_n)) \right| = |nt^{n-1}\psi(t/r_n) + \frac{t^n}{r_n}\psi'(t/r_n)| \leq n r_n^{n-1} + M_1 r_n^{n-1} = (n+M_1)r_n^{n-1},\]
where $M_1 = \sup_{u \in [-1,1]} |\psi'(u)|$. The corresponding series can be shown to converge uniformly.
Therefore, the series (\ref{cutF}) converges to a function $V(t,x)$ that is smooth on $\mathbb{R}\times \Omega$. Verifying that the Taylor series of this function $V(t,x)$ with respect to $t$ at $t=0$ coincides with (\ref{series2}) is straightforward. The proof for $p>1$ and $m>1$ is completely analogous. $\Box$

The formal solution (\ref{Ldsol2}) to the Cauchy problems (\ref{NLdE2}), (\ref{LdE2}), and (\ref{Nonl3}) is not necessarily a smooth function. However, by the Borel-Whitney theorem, there exists a smooth function $\tilde{v}$ of the form
\begin{equation}
\tilde{v}(t,x)=\sum_{n=0}^\infty\frac{t^n A^n u(x)}{n!}\psi(t/r_n),
\label{solf}
\end{equation}
which has the same Taylor expansion at $t=0$ as the formal solution $v(t,x)$. Here, $\psi(t)$ is a cutoff function.

An important remark is in order. The formulation of the Borel-Whitney Lemma implies an ambiguity in the function $\tilde{v}(t,x)$, which is only defined up to a flat function. However, when constructing the solution using expression (\ref{solf}), the coefficients are uniquely determined by known functions ($u$ and $F$). Consequently, there is no ambiguity or additional flat function introduced in this construction. This is consistent with the well-known uniqueness of solutions for linear equations obtained via the Picard iteration method.

Thus, we can formulate the following proposition:

\textbf{Proposition 2.} \emph{Let an operator $A$ of type} (\ref{opA}) \emph{be defined with functions} $u, F \in C^\infty$ \emph{for} $x \in \mathbb{R}^n$. \emph{Then there exists a smooth function} $\tilde{v}(t,x)$ \emph{of the form} (\ref{solf}) \emph{that has the same power series expansion at} $t=0$ \emph{as} $v(t,x)$, \emph{and this function is a solution to equations} (\ref{NLdE2}), (\ref{LdE2}), \emph{and} (\ref{Nonl3}).

\section{Some Generalizations}
Previously, we considered single equations whose right-hand side does not explicitly depend on $t$; consequently, the operator $A$ we introduced was also time-independent. Here, we briefly consider systems of equations and equations whose right-hand side explicitly depends on $t$. In doing so, we rely on the ideas presented in \cite{Kosovtsov}.

Consider a system of $m$ equations for $m$ unknown functions $v_i(t,x)$ ($t,x\in \mathbb{R}\times \mathbb{R}^n$):
\begin{equation}
\frac{\partial v_i}{\partial t} = F_i(x,\textbf{v},\textbf{D}_j^{\alpha_j}\textbf{v}),\qquad i=1,...,m,\quad j\in \mathbb{N}, \qquad v_i(t,x)|_{t=0}=u_i(x),
\label{sysNLdE}
\end{equation}
where $v_i=v_i(t,x)$, $\textbf{v}=(v_1,...,v_m)$, and $\textbf{D}_j^{\alpha_j}\textbf{v}$ is a given finite sequence of derivatives of the functions $v_i$ with respect to the components of $x \in \mathbb{R}^n$.

We now introduce a \emph{single} operator $A$:
\begin{equation}
A=\int_{\mathbb{R}^n} d\zeta \sum_{k=1}^m F_k(\zeta,\textbf{u}(\zeta),\textbf{D}_j^{\alpha_j}\textbf{u}(\zeta))\frac{\delta}{\delta u_k(\zeta)},
\label{sysopA}
\end{equation}
where $u_i=u_i(x)$, $i=1,...,m$, is the set of initial conditions.

This operator $A$ is also time-independent. Therefore, by following essentially the same procedure as in Section 2, we obtain a formal solution to the system (\ref{sysNLdE}):
\begin{equation}
v_i(t,x)=\sum_{n=0}^\infty\frac{t^n A^n}{n!}u_i(x)=e^{tA}u_i(x),\qquad (i=1,...,m).
\label{sysLdsol}
\end{equation}
As a consequence of Propositions 1 and 2, if the functions $u_i, F_i$ are $C^\infty$ for $x \in \mathbb{R}^n$, then there exist smooth functions $\tilde{v}_i(t,x)$ of the form (\ref{solf}) that have the same power series expansions at $t=0$ as $v_i(t,x)$.

If the right-hand sides of the equations explicitly depend on $t$,
\begin{equation}
\frac{\partial v_i}{\partial t} = F_i(t,x,\textbf{v},\textbf{D}_j^{\alpha_j}\textbf{v}),\qquad i=1,...,m,\quad j\in \mathbb{N}, \qquad v_i(t,x)|_{t=0}=u_i(x),
\label{systNLdE}
\end{equation}
then the corresponding operator $A$ is also time-dependent:
\begin{equation}
A(t)=\int_{\mathbb{R}^n} d\zeta \sum_{k=1}^m F_k(t,\zeta,\textbf{u}(\zeta),\textbf{D}_j^{\alpha_j}\textbf{u}(\zeta))\frac{\delta}{\delta u_k(\zeta)}.
\label{systopA}
\end{equation}
In this case, the operators $A(t_1)$ and $A(t_2)$ do not commute for $t_1\neq t_2$, which leads to significant complications, starting with the interpretation of the Picard iteration method. To represent the resulting expansion as an operator exponent, one usually introduces the time-ordered (or chronological) exponent, proposed by F. Dyson, whose properties differ significantly from those of an ordinary operator exponent \cite{Kosovtsov3}. Crucially for our purposes, this approach does not directly yield a solution in the form of a Taylor series expansion in $t$.

In \cite{Kosovtsov}, \cite{Kosovtsov3}, a technique was proposed to reduce the time-ordered exponent to an ordinary one by modifying the operator $A(t)$ (\ref{systopA}). This is achieved by replacing $t\rightarrow s$ and adding the term $\frac{\partial }{\partial s}$:
\begin{equation}
\tilde{A}=\int_{\mathbb{R}^n} d\zeta \sum_{k=1}^m F_k(s,\zeta,\textbf{u}(\zeta),\textbf{D}_j^{\alpha_j}\textbf{u}(\zeta))\frac{\delta}{\delta u_k(\zeta)}+\frac{\partial }{\partial s}.
\label{syssopA}
\end{equation}
The operator $\tilde{A}$ is now independent of $t$, and therefore we obtain a formal solution to the system (\ref{systNLdE}):
\begin{equation}
v_i(t,x)=\left[\sum_{n=0}^\infty\frac{t^n\tilde{A}^n}{n!}u_i(x)\right]_{s=0}=\left[e^{t\tilde{A}}u_i(x)\right]_{s=0},\qquad (i=1,...,m),
\label{syssLdsol}
\end{equation}
namely, in the form of a Taylor series expansion of $v_i(t,x)$ in $t$ at $t=0$. The consequences of Propositions 1 and 2 then follow as before.

This list of possible generalizations is far from complete.

\section{Example: Smoothness of Solutions to the Navier-Stokes Equations}
Incompressible flows of homogeneous fluids in the whole space $\mathbb{R}^3$ are described by the system of equations \cite{Ladyzh}-\cite{Gilles}:
\begin{equation}
\frac{\partial v}{\partial t} = \nu\Delta v - (v \cdot \nabla) v - \nabla p + f,
\label{N-Sv}
\end{equation}
\begin{equation}
\mathrm{div} \, v = \sum_{j=1}^3 \frac{\partial v_j}{\partial x_j}=0, \qquad (x,t) \in \mathbb{R}^3 \times [0,\infty),
\label{N-Svo}
\end{equation}
\begin{equation}
v|_{t=0}=u, \qquad x \in \mathbb{R}^3,
\label{N-V0}
\end{equation}
where $v(x, t) = (v_1, v_2, v_3)$ is the fluid velocity field, $p(x, t)$ is the scalar pressure, $\nabla$ is the gradient operator, and $\Delta \equiv \sum_{j=1}^3 \frac{\partial^2}{\partial x_j^2}$ is the Laplacian. The term $(v \cdot \nabla) v$ is shorthand for $\sum_{j=1}^3 v_j \frac{\partial v}{\partial x_j}$. The constant $\nu \geq 0$ is the kinematic viscosity, $u(x)$ is the initial velocity field, and $f(x,t)$ is a given external force. Henceforth, we will consider the Navier-Stokes equations in the absence of external forces, i.e., we set $f=0$.

We restrict our attention to the Navier-Stokes equations on the whole space $\mathbb{R}^3$, thereby avoiding the delicate issue of boundary conditions. The operator $A$ will be defined accordingly.

It is well known \cite{Majda}, \cite{Foias} that the pressure term can be eliminated by solving for $p$ in terms of $v$. Taking the divergence of both sides of the momentum equation (\ref{N-Sv}) and applying the incompressibility condition (\ref{N-Svo}), we find that
\begin{equation}
-\Delta p = \sum_{i,j=1}^3 \frac{\partial v_j}{\partial x_i} \frac{\partial v_i}{\partial x_j} = \mathrm{div} ((v \cdot \nabla)v).
\label{eqPoisson}
\end{equation}
Equation (\ref{eqPoisson}) is the well-known Poisson equation for the pressure $p$. Its solution in $\mathbb{R}^3$ is given by \cite{Majda} (for brevity, we denote this solution as $p_v(x,t)$ or simply $p_v$):
\begin{equation}
p_v(x,t) = \frac{1}{4\pi} \int_{\mathbb{R}^3} \sum_{i,j=1}^3 \frac{\partial v_j(t,\xi)}{\partial \xi_i} \frac{\partial v_i(t,\xi)}{\partial \xi_j} \frac{1}{|x-\xi|} \, d\xi.
\label{Poissonsoln}
\end{equation}
Substituting this solution (\ref{Poissonsoln}) back into (\ref{N-Sv}), we obtain the Leray formulation of the Navier-Stokes equations, a \emph{nonlinear system of integro-differential equations} for the velocity field $v$:
\begin{equation}
\frac{\partial v}{\partial t} = \nu\Delta v - (v \cdot \nabla) v - \nabla p_v, \qquad (x \in \mathbb{R}^3, t \geq 0).
\label{N-Svv}
\end{equation}
It is well-established \cite{Majda} that the original Navier-Stokes system (\ref{N-Sv})-(\ref{N-V0}) is equivalent to the integro-differential system (\ref{N-Svv}) with the pressure defined by (\ref{Poissonsoln}), provided that the initial data $u(x)$ is smooth and satisfies the divergence-free condition:
\begin{equation}
\sum_{i=1}^3 \frac{\partial u_i}{\partial x_i}=0, \qquad (x \in \mathbb{R}^3).
\label{N-S20}
\end{equation}

Although the Leray formulation has been known for a long time, it has not been widely used to analyze solutions. The main reasons for this are as follows \cite{Majda}:
The equation (\ref{N-Svv}) is quadratically nonlinear and contains a nonlocal, quadratically nonlinear operator. These facts make the Navier-Stokes equations difficult to study analytically.
Although (\ref{N-Svv}) constitutes a closed system for $v$, this formulation is of limited use for detailed analysis, apart from methods based on the energy principle. The main reason is that this formulation obscures properties related to vorticity dynamics, such as vortex stretching.

Let the initial velocity field $u(x) = (u_1(x), u_2(x), u_3(x))$ be a smooth vector function, i.e., $u_i \in C^\infty(\mathbb{R}^3)$ for each component. In accordance with the framework developed previously, we introduce the following \textit{linear} differential operator:
\begin{align}
A = \int_{\mathbb{R}^3} \sum_{i=1}^3 &\biggl\{ \nu\Delta_{\zeta} u_i(\zeta)  - \sum_{j=1}^3 u_j(\zeta) \frac{\partial u_i(\zeta)}{\partial \zeta_j} \notag \\
& - \frac{1}{4\pi}\frac{\partial}{\partial \zeta_i} \biggl[ \int_{\mathbb{R}^3} \sum_{k,l=1}^3 \frac{\partial u_l(\xi)}{\partial \xi_k} \frac{\partial u_k(\xi)}{\partial \xi_l} \frac{1}{|\zeta-\xi|} \, d\xi \biggr] \biggr\} \frac{\delta}{\delta u_i(\zeta)} \, d\zeta.
\label{OPERn}
\end{align}
In a more compact notation:
\begin{equation}
A=\int_{\mathbb{R}^3} \sum_{i=1}^3 \left\{ \nu\Delta_{\zeta} u_i(\zeta) - \sum_{j=1}^3 u_j(\zeta) \frac{\partial u_i(\zeta)}{\partial \zeta_j} - \frac{\partial p_u(\zeta)}{\partial \zeta_i} \right\} \frac{\delta}{\delta u_i(\zeta)} \, d\zeta,
\label{OPERp}
\end{equation}
where
\begin{equation}
p_u(\zeta) \equiv \frac{1}{4\pi} \int_{\mathbb{R}^3} \sum_{k,l=1}^3 \frac{\partial u_l(\xi)}{\partial \xi_k} \frac{\partial u_k(\xi)}{\partial \xi_l} \frac{1}{|\zeta-\xi|} \, d\xi.
\label{up}
\end{equation}
It is well known that if $u_i \in C^\infty$, then $p_u$ is also $C^\infty$. Therefore, the operator $A$ satisfies the conditions of Propositions 1 and 2.

Consequently, there exist smooth functions $\tilde{v}_i(t,x)$ of the form (\ref{solf}) that have the same power series expansions at $t=0$ as the formal solutions $v_i(t,x)$. These functions solve equation (\ref{N-Svv}) and, accordingly, the original Navier-Stokes system (\ref{N-Sv}), provided that the initial data $u_i$ are smooth and satisfy the condition (\ref{N-S20}).

Note that the constructive scheme presented here is also applicable for $x \in \mathbb{R}^n$ for any $n \in \mathbb{N}$ and in the presence of smooth external forces. Furthermore, setting the kinematic viscosity $\nu=0$ allows these results to be directly applied to the Euler equations.


\begin{thebibliography}{9}

\bibitem {Kosovtsov} Yu. N. Kosovtsov: Formal exact operator solutions to nonlinear differential equations,  Preprint at arXiv: \url{https://arxiv.org/abs/0910.3923},  (2009).

\bibitem {Kosovtsov3} Yu. N. Kosovtsov: The Chronological Operator Algebra and Formal Solutions of Differential Equations, Preprint at arXiv: \url{https://arxiv.org/abs/math-ph/0409035},  (2004).

\bibitem {Kosovtsov2} Yu. N. Kosovtsov: Formal series solutions to nonlinear DE (ODE or PDE) or systems of them (Cauchy problem), Application, Maplesoft Application Center, (2005),  \url{ https://www.maplesoft.com/applications/Detail.aspx?id=1690}.
\bibitem {Poston}  T. Poston, I. Stewart: Catastrophe Theory and Its Applications, Pitman, (1979).

\bibitem {Borel} E. Borel:  Sur quelques points de la th\'{e}orie des fonctiones, Ann. Sci. \'{E}cole Normal Superior 3,  p 9-55, (1895).

\bibitem {Whitney1} H. Whitney:  Analytic extensions of differentiable functions defined in closed sets, Trans. Amer. Math. Soc. 36,  p 63-89, (1934).
\bibitem {Whitney2}  H. Whitney: Differentiable functions defined in closed sets, I, Trans. Amer. Math. Soc. 36,  p 369-387,  (1934).

\bibitem {Brudnyi} A. Brudnyi, Yu. Brudnyi:  Methods of Geometric Analysis in Extension and Trace Problems, Volumes 1-2, Birkh\"{a}user Basel,  (2012),  \url{https://doi.org/10.1007/978-3-0348-0209-3}.

\bibitem {Fefferman} Ch. Fefferman: A sharp form of Whitney's extension theorem, Ann. Math. 161, p 509-577, (2005), \url{https://doi.org/10.4007/annals.2005.161.509}.

\bibitem {Golubitsky} M. Golubitsky, V. Guillemin: Stable Mappings and Their Singularities, Springer-Verlag, New York, (1973).
\bibitem {Hormander}  L. H\"{o}rmander: The Analysis of Linear Partial Differential Operators. I, Distribution Theory and Fourier Analysis, Springer-Verlag, (1973).

\bibitem {Ladyzh} O. Ladyzhenskaya:  The Mathematical Theory of Viscous Incompressible Flows (2nd edition),Gordon and Breach, New York, (1969).
\bibitem {NAV} Ch. Fefferman: Existence and smoothness of the Navier-Stokes equation, (2000), \url{ http://claymath.org/Millenium-Prize-Problems/Navier-Stokes-Equations}, Clay
Mathematics Institute, Cambridge, MA.
	\bibitem {Gilles} P. G. Lemari\'{e}-Rieusset: The Navier-Stokes Problem in the 21st Century, Chapman and Hall, (2020), \url{https://doi.org/10.1201/9781315373393}.
\bibitem {Majda} A. J. Majda, A. L. Bertozz: Vorticity and incompressible flow, Cambridge University Press,(2001),  \url{https://doi.org/10.1115/1.1483363}.
\bibitem {Foias} C. Foias, O. Manley, R. Rosa, R. Temam : Navier-Stokes equations and turbulence (Encyclopedia of mathematics and its applications 83), Cambridge University Press ,  (2001), \url{https://doi.org/10.1017/CBO9780511546754}.

\end{thebibliography}
\end{document}